\documentclass[preprint,12pt]{elsarticle}

\usepackage{graphicx}
\usepackage{subfigure}

\usepackage{amssymb}

\journal{Solid State Communications}

\begin{document}

\begin{frontmatter}



\title{
Chemical Trend of Superconducting Transition Temperature
 in Hole-doped Delafossite of CuAlO$_2$, AgAlO$_2$ and AuAlO$_2$ 
}

\author{Akitaka Nakanishi\corref{cor}}
\ead{nakanishi@aquarius.mp.es.osaka-u.ac.jp}
\author{Hiroshi Katayama-Yoshida}
\cortext[cor]{Tel.:+81-6-6850-6504. Fax:+81-6-6850-6407.}
\address{Department of Materials of Engineering Science,
 Osaka University, Toyonaka, Osaka 560-8531, Japan}

\begin{abstract}
We have performed the first-principles calculations about
 the superconducting transition temperature $T_{\rm c}$ of
 hole-doped delafossite CuAlO$_2$, AgAlO$_2$ and AuAlO$_2$.
Calculated $T_{\rm c}$ are about 50 K(CuAlO$_2$), 40 K(AgAlO$_2$)
 and 3 K(AuAlO$_2$) at maximum in the optimum hole-doping concentration.
The low $T_{\rm c}$ of AuAlO$_2$ is attributed to
 the weak electron-phonon interaction caused by the low covalency
 and heavy atomic mass.
\end{abstract}


\end{frontmatter}


\section{Introduction}
CuAlO$_2$ has a delafossite structure (Fig. \ref{fig:str})
 and a two-dimensional electronic structure
 caused by the natural super-lattices of O-Cu-O dumbbell.
Kawazoe {\it et al.} have found that
 the CuAlO$_2$ is $p$-type transparent
 conducting oxides (TCO) without any intentional doping.\cite{Kawazoe1997}
Nakanishi {\it et al.} studied the pressure dependence of the structures
 \cite{Nakanishi2011a} and the role of the self-interaction correction
 in CuAlO$_2$.\cite{Nakanishi2011b}
Transparent $p$-type conductors such as CuAlO$_2$ are
 important for the $p$-$n$ junction of TCO
 and a realization of high-efficiency photovoltaic solar-cells.
First-principles calculations have shown the possibility
 of applications for high efficiency thermoelectric power with about 1\% hole-doping.
 \cite{Funashima2004,Yoshida2009,Hamada2006}

Katayama-Yoshida {\it et al.} have simulated
 the Fermi surface of the $p$-type doped CuAlO$_2$
 by shifting the Fermi level rigidly
 and proposed that
 the nesting Fermi surface may cause
 a strong electron-phonon interaction
 and a transparent superconductivity for visible light. \cite{Yoshida2003}
But,
 the calculation of superconducting transition temperature $T_{\rm c}$
 was not carried out.
In previous study,
 we calculated the $T_{\rm c}$ of $p$-type doped CuAlO$_2$ \cite{Nakanishi2012}
 and found that the $T_{\rm c}$ goes up to about 50\,K
 due to the strong electron-phonon interaction
 by the two dimensional flat valence band.
The origin of the flat band is the $\pi$-band of
 hybridized O 2p$_z$ and Cu 3d$_{3z^2-r^2}$
 on the frustrated triangular lattice in the two dimensional plane.
It is very interesting to see the relation between the $T_{\rm c}$
 and the flatness of the flat band by changing the Cu 3d$_{3z^2-r^2}$,
 Ag 4d$_{3z^2-r^2}$ and Au 5d$_{3z^2-r^2}$.
In this study,
 we performed the calculation of $T_{\rm c}$ and the electron-phonon interaction
 about the chemical trend of $p$-type doped CuAlO$_2$, AgAlO$_2$ and AuAlO$_2$.

\section{Calculation Methods}
The calculations were performed
 within the density functional theory\cite{Hohenberg1964,Kohn1965}
 with a plane-wave pseudopotential method,
 as implemented in the Quantum-ESPRESSO code.\cite{Giannozzi2009}
We employed the Perdew-Wang 91
 generalized gradient approximation (GGA)
 exchange-correlation functional\cite{Perdew1992}
 and ultra-soft pseudopotentials.\cite{Vanderbilt1990}
For the pseudopotentials,
 3d electrons of transition metals were also included
 in the valence electrons.
In reciprocal lattice space integral calculation,
 we used $8\times8\times8$ (electron and phonon)
 and $32\times32\times32$ (density of states and average at Fermi level)
  ${\bf k}$-point grids
 in the Monkhorst-Pack grid.\cite{Monkhorst1976}
The energy cut-off for wave function was 40  Ry and
 that for charge density was 320  Ry.
These ${\bf k}$-point meshes are fine
 enough to achieve convergence
 within 10 mRy/atom in the total energy.

\begin{figure}[htbp]
  \begin{center}
    \includegraphics{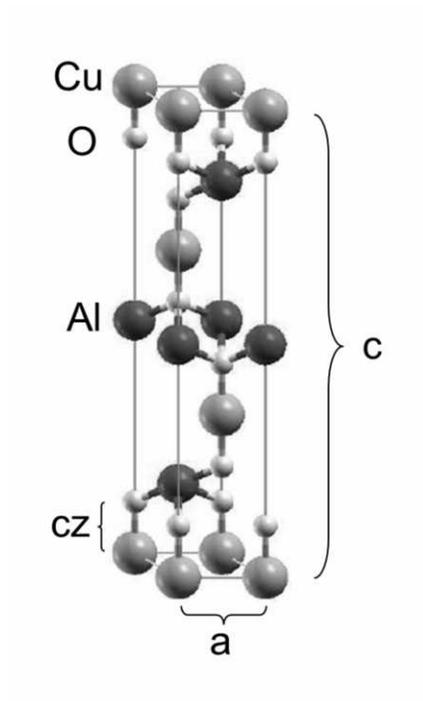}
    \caption{The crystal structure of delafossite CuAlO$_2$.}
    \label{fig:str}
  \end{center}
\end{figure}

The delafossite structure belongs to the space group R$\bar3$m (No.166)
 and is represented by cell parameters $a$ and $c$,
 and internal parameter $z$ (See Fig. \ref{fig:str}).
These cell parameters and internal parameter were optimized by 
 the constant-pressure variable-cell relaxation using
 the Parrinello-Rahman method\cite{Parrinello1980}
 without any symmetry requirements.

In this study,
 some properties of hole-doped materials are approximated
 because it is difficult for first-principles calculation
 to deal with the doped system exactly.
Let's take the electron-phonon interaction $\lambda$ for example.
$\lambda$ is defined as follows:

\begin{equation}
 \lambda
 =
 \sum_{\nu{\bf q}}
 \frac{
 2N(\varepsilon_{\rm F})
 \sum_{{\bf k}}|M_{{\bf k,k+q}}^{\nu{\bf q}}|^2
 \delta(\varepsilon_{\bf k}-\varepsilon_{\rm F})
 \delta(\varepsilon_{{\bf k+q}}-\varepsilon_{\rm F})
 }
 {
 \omega_{\nu{\bf q}}\sum_{{\bf kq'}}
 \delta(\varepsilon_{\bf k}-\varepsilon_{\rm F})
 \delta(\varepsilon_{{\bf k+q'}}-\varepsilon_{\rm F})
 }.
\label{eqn:lambda}
\end{equation}
(1)
 For the non-doped systems,
 we calculated the dynamical matrix,
 the phonon frequency $\omega_{\nu{\bf q}}$ 
 and the electron-phonon matrix $M_{{\bf k,k+q}}^{\nu{\bf q}}$.
(2)
 For the doped systems,
 we calculated the Fermi level $\varepsilon_{\rm F}$ and
 the density of states at the Fermi level $N(\varepsilon_{\rm F})$
 with the number of valence electrons reduced
 using the eigenvalues $\varepsilon_{\bf k}$ of the non-doped system.
(3)
 By using the results of (1) and (2),
 we calculated the electron-phonon interaction $\lambda$
 and the other superconducting properties.
This approximation is based on
 the idea that the doping does not greatly change
 electron and phonon band structures.
In this study,
 we show the results of $0.1\sim1.0$ hole-doped systems.

We calculated the superconducting transition temperature
 by using the Allen-Dynes modified McMillan's formula.\cite{McMillan1968,Allen1975} 
According to this formula,
 $T_{\rm c}$ is given by three parameters:
 the electron-phonon interaction $\lambda$, 
 the logarithmic averaged phonon frequency $\omega_{\log}$, 
 and the screened Coulomb interaction $\mu^{\ast}$, in the following form.
\begin{eqnarray}
  T_{\rm c}&=&\frac{\omega_{\log}}{1.2}
  \exp \left( \frac{-1.04(1+\lambda )}
  {\lambda-\mu^{\ast}(1+0.62\lambda )} \right). \\
  \omega_{\log} 
  &=&\exp \left(\frac{2}{\lambda}\int_0^\infty
  d\omega\frac{\alpha^2F(\omega)}{\omega}\log\omega\right).
\end{eqnarray}
Here, $\alpha^2F(\omega)$ is the Eliashberg function.
$\lambda$ and $\omega_{\rm log}$ are obtained by 
 the first-principle calculations using
 the density functional perturbation theory.\cite{Baroni2001}
As for $\mu^{\ast}$,
 we assume the value $\mu^{\ast}=0.1$.
This value holds for weakly correlated materials.

\section{Calculation Results and Discussion}
Before superconducting calculation,
 we optimized the cell parameters.
The optimization results show that
 the structural transition does not occur
 and the delafossite structure is stable.
Table 1 
 shows the optimized cell parameters.
The distance between transition metal and oxygen is the longest in CuAlO$_2$.
This shows the strong covalency of CuAlO$_2$.
The calculation results of CuAlO$_2$ agree very well with the experimental data
 ($a=2.858$\AA, $c/a=5.934$ and $z=0.1099$ \cite{Nie2002,Buljan1999}).

\begin{table} [htpd]
\begin{center}
  \begin{tabular}{cccc}
  \hline
            & CuAlO$_2$ & AgAlO$_2$ & AuAlO$_2$ \\ \hline 
  $a$ [\AA] & 2.859     & 2.895     & 2.913     \\
  $c/a$     & 5.965     & 6.369     & 6.373     \\
  $z$       & 0.1101    & 0.1148    & 0.1158    \\
  \hline
  \end{tabular}
  \label{tab:opt}
  \caption{The optimized cell parameters.}
\end{center}
\end{table}

\begin{figure}[htbp]
\begin{center}
  \includegraphics{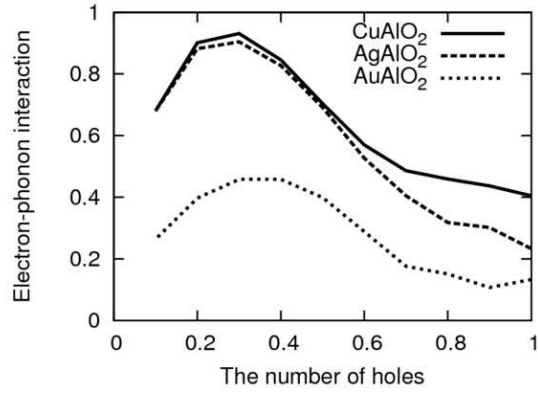}
  \caption{Electron-phonon interaction $\lambda$.}
  \label{fig:lambda}
\end{center}
\end{figure}

Fig. \ref{fig:lambda} shows the calculated results of
 electron-phonon interaction $\lambda$.
AuAlO$_2$ has lower $\lambda$ than CuAlO$_2$ and AgAlO$_2$.
On the other hand, 
 $\lambda$ of AgAlO$_2$ is close to that of CuAlO$_2$.
This difference is attributed to not only atomic mass but also covalency.

\begin{figure}[htbp]
\begin{center}
  \includegraphics{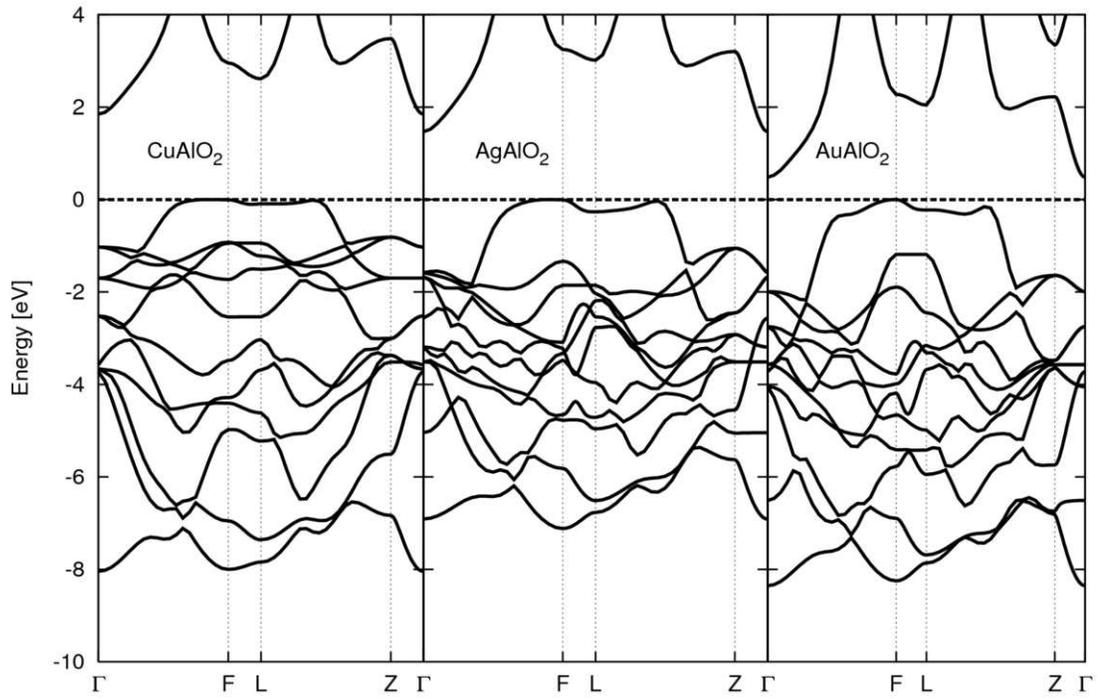}
  \caption{Band structures of CuAlO$_2$(left), AgAlO$_2$(center) and AuAlO$_2$(right).}
  \label{fig:band}
\end{center}
\end{figure}

\begin{figure}[htbp]
\begin{center}
  \includegraphics{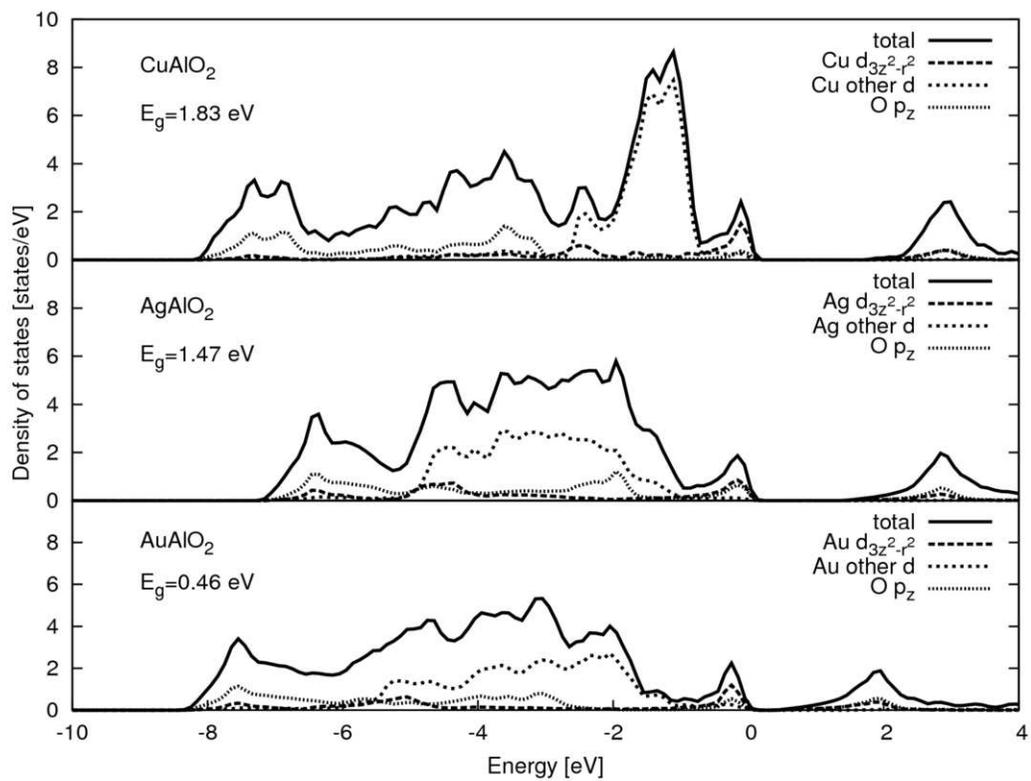}
  \caption{Density of states and energy gaps.}
  \label{fig:dos}
\end{center}
\end{figure}

Fig. \ref{fig:band} and \ref{fig:dos} show the band structure, the density of states and energy gaps.
The energy gaps are 1.83eV(CuAlO$_2$), 1.47(AgAlO$_2$) and 0.46(AuAlO$_2$). 
As $d$-wave function extends ($3d\rightarrow4d\rightarrow5d$),
 the $d$-band widen and the covalent bonding weaken.
So this low covalency of AuAlO$_2$ is the cause of his low electron-phonon interaction $\lambda$.
It is also clearly seen in Fig. \ref{fig:band} that the flatness of
 the top of the valence band decreases from CuAlO$_2$ to AuAlO$_2$
 with reducing the covalency.

\begin{figure}[htbp]
\begin{center}
  \includegraphics{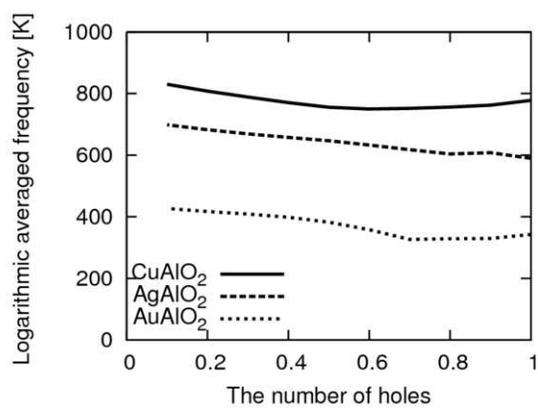}
  \caption{Logarithmic averaged phonon frequency $\omega_{\log}$.}
  \label{fig:omega}
\end{center}
\end{figure}

Fig. \ref{fig:omega} shows the calculated results of
 logarithmic averaged phonon frequency $\omega_{\log}$.
These are almost constant for the number of holes.
CuAlO$_2$ has the highest $\omega_{\log}$.
The difference of $\omega_{\log}$ is due to atomic mass of transition metals.

\begin{figure}[htbp]
\begin{center}
  \includegraphics{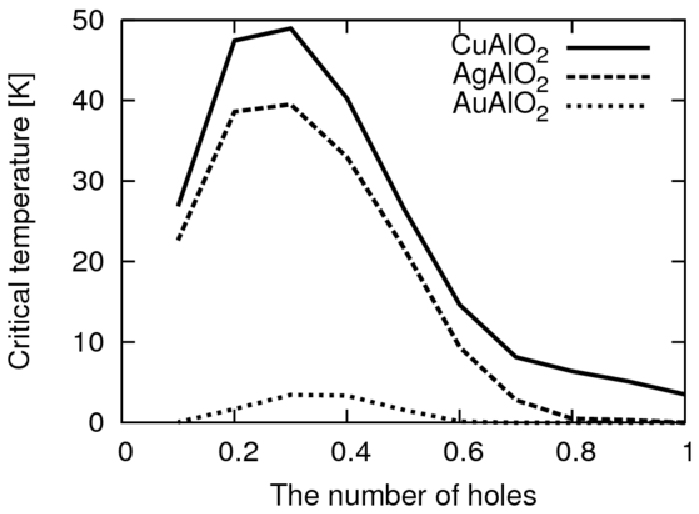}
  \caption{Superconducting transition temperature $T_{\rm c}$.}
  \label{fig:tc}
\end{center}
\end{figure}

Fig. \ref{fig:tc} shows the calculated results of $T_{\rm c}$.
The lightly hole-doped material ($N_{\rm h}=0.2\sim0.3$) has higher $T_{\rm c}$
 while the heavily doped material ($N_{\rm h}=0.6\sim1.0$) has lower $T_{\rm c}$.
AgAlO$_2$ has a little lower $T_{\rm c}$ than CuAlO$_2$.
On the other hand, AuAlO$_2$ has much lower $T_{\rm c}$ than these
 due to the heavy atomic mass and the low covalency mentioned above.

\section{Conclusions}
In summary,
 we calculated the chemical trend of superconducting transition temperature of
 the hole-doped delafossite CuAlO$_2$, AgAlO$_2$ and AuAlO$_2$.
Calculated $T_{\rm c}$ are about 50 K(CuAlO$_2$), 40 K(AgAlO$_2$) and 3 K(AuAlO$_2$) at maximum.
The low $T_{\rm c}$ of AuAlO$_2$ is attributed to low covalency and heavy atomic mass.

\section*{Acknowledgment}
The authors acknowledge the financial support from
 the Global Center of Excellence (COE) program "Core Research and
 Engineering of Advanced Materials - Interdisciplinary Education Center for
 Materials Science", the Ministry of Education, Culture, Sports, Science and
 Technology, Japan, and a Grant-in-Aid for Scientific Research on Innovative
 Areas "Materials Design through Computics: Correlation and Non-Equilibrium Dynamics".
We also thank to the financial support from the Advanced Low Carbon Technology Research and 
Development Program, the Japan Science and Technology Agency for the financial support.
\bibliographystyle{elsarticle-num}
\bibliography{bibfile}







\end{document}